\documentstyle[sprocl]{article}
\renewcommand{\theequation}{\arabic{section}.\arabic{equation}}

\input{psfig.sty}

\def\be{\begin{equation}}
\def\ee{\end{equation}}
\def\bea{\begin{eqnarray}}
\def\eea{\end{eqnarray}}
\def\ba{\begin{array}}
\def\ea{\end{array}}
\def\psibar{\bar{\psi}}
\def\zetabar{\bar{\zeta}}
\def\lambdabar{\bar{\lambda}}
\def\Ap2{A(p^2)}
\def\Aq2{A(q^2)}
\def\Bp2{B(p^2)}
\def\Bq2{B(q^2)}
\def\slash{\gamma \cdot}
\def\half{\frac{1}{2}}
\def\Tr{\mbox{Tr}}

\begin{document}
\setlength{\unitlength}{1mm}

\title{NONPERTURBATIVE VERTICES IN SUPERSYMMETRIC QUANTUM ELECTRODYNAMICS}

\author{M. L. WALKER and C. J. BURDEN}

\address{Department of Theoretical Physics,\\ Research School of 
Physical Sciences and 
Engineering,\\ Australian National University, Canberra, A.C.T. 0200
Australia\\E-mail: mlw105@rsphysse.anu.edu.au, conrad.burden@anu.edu.au}

\maketitle

\begin{abstract}
We derive the complete set of supersymmetric Ward identities involving
only two- and three- point proper vertices in supersymmetric QED. We also 
present the most general form of the proper vertices consistent with both the
supersymmetric and $U(1)$ gauge Ward identities. These vertices are the 
supersymmetric equivalent of the non supersymmetric Ball-Chiu vertices.
\end{abstract}

\setcounter{equation}{0}
\section{Introduction}

While supersymmetry (SUSY) is generally agreed to be integral to any
theory incorporating both gravity and gauge forces, 
techniques for investigating nonperturbative effects such as 
chiral symmetry breaking are still in their early stages. 
A small number of authors
have employed Dyson Schwinger Equations (DSEs) to analyse various SUSY theories
\cite{Pski,clark,ster,sham} and small inroads into numerical solutions 
\cite{me} of the SUSY DSE (SDSE) in Supersymmetric Quantum Electrodynamics 
(SQED) in 2+1 dimensions (SQED$_3$) have been made. 

Analyses of SUSY theories generally use the rainbow approximation to 
truncate the DSEs at a manageable level. One exception is Clark and Love who 
use the superfield formalism and
derive a differential $U(1)$ gauge Ward Identity for the superfields. 
They find that the effective mass contains a prefactor 
which vanishes in Feynman gauge and conclude that there can be no 
spontaneous mass generation in SQED, even beyond the rainbow approximation.
However the superfield 
approach suffers the disadvantage that each DSE contains an infinite 
number of terms.  This is dealt with by truncating
diagrams containing seagull and higher order $n$-point vertices.  

The work of Clark and Love has been criticized by 
Kaiser and Selipsky on two grounds \cite{kaiser}.  Firstly they argue that 
the truncation of seagull diagrams is too severe as it ignores contributions 
even at the one-loop level.  Secondly they point out that infinities 
arising from infrared divergences which plague the superfield formalism 
can counter the vanishing prefactor and allow spontaneous mass generation.  
These criticisms highlight some of the dangers of attempting to extract 
phenomenological consequences of supersymmetric DSEs by working 
solely with the superfield formalism.  In fact, analyses in the literature
\cite{sham,me} have generally found the component formalism to be the 
most efficient way to proceed.  

Koopmans and Steringa \cite{ster}, using the 
component formalism, also sought to be consistent with the
differential $U(1)$ gauge WI in their analysis of SQED3 with two-component
fermions. To this end they
multiplied the bare vertices by $\Aq2$ where the electron propagator is given
by $S^{-1}(q)=i(\slash q A(q^2) + B(q^2))$. This approach is questionable
as it implicitly approximates the functions $A(p^2)$ and $B(p^2)$ as being
flat. While this approximation is reasonable 
over most of the momentum range, it
is not valid in the low momentum limit where the dynamics are
largely determined.

Attempts to go beyond the rainbow approximation in non-SUSY theories 
began with the Ball and Chiu \cite{ball} vertex ans\"atze
for QED and QCD. These are the minimal vertices which ``solve'' the Ward 
Takahashi Identities (WTIs) while avoiding kinematic
singularities. Ball and Chiu also gave the general form of the possible 
``transverse'' pieces which may be added. Since then several authors have 
sought to construct ans\"atze which improve on the minimal Ball-Chiu
vertex \cite{CP91,paulus}.

That analogous progress has not been made in SQED using the component formalism
is not suprising. Not only must the gauge particle vertices be dressed
but the gaugino vertices also. Indeed substituting the minimal Ball and Chiu
vertex for photon interactions in SQED$_3$ while leaving the other vertices
bare exacerbates the SDSE's gauge violating properties \cite{me}. The problem 
of going beyond the rainbow approximation in SUSY theories is the problem of
finding the gaugino vertices corresponding to the improved photon vertex. 
Gaugino vertices are not constrained by the WTI since the gaugino
is invariant to gauge transformations. However they are related to the gauge
particle vertices by SUSY Ward Identities (SWIs). It is the purpose of this 
paper to derive and solve the SWIs for SQED and obtain the most general
form of the three-point vertex functions consistent with both SUSY and $U(1)$
gauge Ward identities.

Sec.\ref{twopoint} gives the SWIs between the various two-point functions
of SQED and their solution which is unique once the electron
propagator is known. Sec.\ref{auxiliary} shows how to
treat proper functions of auxiliary fields. 
Sec.\ref{SWI} gives the SWIs 
constraining the three-point proper functions and finds that
the rainbow approximation violates SUSY. The most general form 
of the vertices consistent with
these identities is presented in Sec.\ref{threepoint} and proven
to be so in Appendix \ref{proof}.

\setcounter{equation}{0}
\section{$U(1)$ and Supersymmetric Ward Identities}
\label{twopoint}

The conventions used in this paper are that $g^{\mu \nu} =$
diag$(1,-1,-1,-1)$, $\{\gamma^\mu,\gamma^\nu\}=2g^{\mu \nu}$, 
and $\gamma_5 = i\gamma^0 \gamma^1 \gamma^2 \gamma^3$.

The Lagrangian of SQED,
\bea \label{lagrang}
L &=& |f|^2 + |g|^2+ |\partial_\mu a|^2 + |\partial_\mu b|^2
- \bar{\psi} \not \! \partial \psi \nonumber \\ \nonumber \\
&&-m(a^*f + af^* + b^*g + bg^* + i\bar{\psi} \psi)\nonumber \\ \nonumber \\
&&- ieA^\mu(a^\ast \stackrel{\leftrightarrow}{\partial}_\mu a
+ b^\ast \stackrel{\leftrightarrow}{\partial}_\mu b
+ \bar{\psi} \gamma_\mu \psi) \nonumber \\ \\
&&- e[\bar{\lambda}(a^\ast + i\gamma_5 b^\ast)\psi 
- \bar{\psi}(a + i\gamma_5 b) \lambda] \nonumber \\ \nonumber \\
&&+ ieD(a^\ast b - a b^\ast)
+e^2 A_\mu A^\mu (|a|^2 + |b|^2) \nonumber \\ \nonumber \\
&&-\frac{1}{4}F^{\mu \nu}F_{\mu \nu} - 
\frac{1}{2}\bar{\lambda} \not \! \partial \lambda + \frac{1}{2} D^2, \nonumber 
\eea
is, by construction, invariant with respect to both $U(1)$
gauge transformations and SUSY transformations where the SUSY transformations
are given by \cite{wz}
\bea \label{super}
&\delta_S a = -i\zetabar \psi,& \nonumber \\
&\delta_S b = \zetabar \gamma_5 \psi,& \nonumber \\
&\delta_S \psi = [f + i\gamma_5 g + i\slash \partial(a + i\gamma_5 b)
	-e \slash A (a-i\gamma_5 b)]\zeta,& \\
&\delta_S f = \zetabar [\slash \partial \psi 
+ e[-a\lambda -i b \gamma_5 \lambda + i\slash A \psi ]],& \nonumber \\
&\delta_S g = i\zetabar [\gamma_5 \slash \partial \psi 
+ e[-\gamma_5 \lambda -i b \lambda -i \slash A \gamma_5 \psi]],& \nonumber
\eea
for the chiral multiplet and
\bea
&\delta_S A_\mu = \zetabar \gamma_\mu \lambda,& \nonumber \\
&\delta_S \lambda = \sigma^{\nu \mu} \partial_\mu A_\nu \zeta
	+i\gamma_5 D \zeta,& \nonumber \\
&\delta_S D = i\zetabar \gamma_5 \slash \partial \lambda,& \nonumber 
\eea
for the vector multiplet. It is important to note that the transformations in
Eqn.(\ref{super}) are not true SUSY transformations but SUSY transformations
plus a gauge transformation. This is a manifestation of the Wess-Zumino (WZ)
gauge which is used to make the Lagrangian polynomial \cite{wz}. A true SUSY 
transformation spoils the WZ gauge and must be followed by a gauge 
transformation which restores it for the Lagrangian to be invariant.
It is from this invariance that the SWIs arise.

The SWIs completely specify the selectron propagators in terms of the electron 
propagator \cite{iz}. The
SWIs relating the scalar propagators to the electron propagator \cite{iz} are 
\begin{equation} \label{SUSY_WI}
\langle \psi \bar{\psi} \rangle  = 
i\langle a^* f\rangle  - i\slash p \langle a^* a\rangle  =
i\langle b^* g\rangle  - i\slash p \langle b^* b\rangle,
\end{equation}
and
\be
\slash p \langle \psi \bar{\psi} \rangle  = 
-i\langle f^* f\rangle  + i\slash p \langle f^* a\rangle  =
-i\langle g^* g\rangle  + i\slash p \langle g^* b\rangle.
\end{equation}
Substituting in the fermion propagator
\begin{equation} \label{fermiprop}
S(p) \equiv \langle \psi \bar{\psi} \rangle  = 
\frac{-i}{\slash p A(p^2) + B(p^2)},
\end{equation}
gives the scalar propagators 
\begin{equation} 
\label{boseprop} 
D(p^2) \equiv \langle a^* a\rangle  = \langle b^* b\rangle  = 
\frac{A(p^2)}{p^2 A^2 (p^2) - B^2 (p^2)}, 
\end{equation}
\begin{equation}
\label{atofprop}
\langle a^* f\rangle  = \langle b^* g\rangle  = 
\frac{B(p^2)}{p^2 A^2 (p^2) - B^2 (p^2)}
= \frac{B(p^2)}{A(p^2)} D(p^2),
\end{equation}
and
\begin{equation} 
\label{auxprop} 
\langle f^* f\rangle  = \langle g^* g\rangle  = 
\frac{p^2 A(p^2)}{p^2 A^2 (p^2) - B^2 (p^2)}.
\end{equation}

SWIs hold between proper vertices too of course. Taking $\Gamma$ to be the
effective action we define $\Gamma_{X..Z} \equiv \frac{\delta^n \Gamma}
{\delta X ... \delta Z}$. The two-point proper vertices are constrained by
\bea
\Gamma_{\psibar \psi} & \equiv & \langle \psi \psibar \rangle^{-1} \nonumber \\
&=& -i\Gamma_{f^* a}(p) + i\slash p \Gamma_{f^* f}(p) =
-i\Gamma_{g^* b}(p) + i\slash p \Gamma_{g^* g}(p),
\eea
\be
\slash p \Gamma_{\psibar \psi}(p) =
i\Gamma_{a^* a}(p) - i\slash p \Gamma_{a^* f}(p) =
i\Gamma_{b^* b}(p) - i\slash p \Gamma_{b^* g}(p),
\ee
to be
\bea \label{twoeleven}
\Gamma_{a^* a}(p) = \Gamma_{b^* b}(p) = p^2 \Ap2, \\
\Gamma_{a^* f}(p) = \Gamma_{f^* a}(p) = \Gamma_{b^* g}(p) =
\Gamma_{g^* b}(p) = - \Bp2, \label{twotwelve} \\
\Gamma_{f^* f}(p) = \Gamma_{g^* g}(p) = \Ap2. \label{twothirteen}
\eea
It is interesting that $\Gamma_{a^* a}(p) = \Gamma_{b^* b}(p) \neq 
D(p^2)^{-1}$. This can be attributed to the presence of the auxiliary
fields $f$ and $g$. The treatment of proper functions involving selectrons
is discussed in the next section.

\setcounter{equation}{0}
\section{Handling the Proper Functions of Auxiliary Fields}
\label{auxiliary}

One of the difficulties of the component notation in SQED is that of 
dealing with the auxiliary fields $f,g$ and $D$. The first two are particularly
difficult as they contribute off-diagonal quadratic terms which 
give the scalar propagators an unfamiliar form. To make the free field
theory manifestly Gaussian we define,
\bea
\mbox{$[a]$} &\equiv& \left( \ba{c} a \\ f \ea \right), \\
\mbox{$[b]$} &\equiv& \left( \ba{c} b \\ g \ea \right), \\
\mbox{$[a]^\dagger$} &\equiv& \left( \ba{cc} a^* & f^* \ea \right), \\
\mbox{$[b]^\dagger$} &\equiv& \left( \ba{cc} b^* & g^* \ea \right).
\eea
The Lagrangian becomes
\bea \label{newlagrang}
L &=& [a]^\dagger \left[ \ba{cc} -\partial^2 & -m \\ -m & 1 \ea \right] [a]
+ [b]^\dagger \left[ \ba{cc} -\partial^2 & -m \\ -m & 1 \ea \right] [b]
- \bar{\psi} (\not \! \partial + im)\psi \nonumber \\ \nonumber \\
&&- ieA^\mu(\, [a]^\dagger \left[ \ba{cc} 
\stackrel{\leftrightarrow}{\partial}_\mu & 0 \\ 0 & 0 \ea \right] [a]
+ [b]^\dagger \left[ \ba{cc} 
\stackrel{\leftrightarrow}{\partial}_\mu & 0 \\ 0 & 0 \ea \right] [b]
+ \bar{\psi} \gamma_\mu \psi)
\nonumber \\ \nonumber \\
&&- e[\bar{\lambda}([a]^\dagger + i\gamma_5 [b]^\dagger)
\left[ \ba{c} 1 \\ 0 \ea \right] \psi 
- \bar{\psi} [ \ba{cc} 1 & 0 \ea ]
([a] + i\gamma_5 [b]) \lambda] \nonumber \\ \\
&&+ ieD([a]^\dagger \left[ \ba{cc} 1 & 0 \\ 0 & 0 \ea \right] [b] 
- [b]^\dagger \left[ \ba{cc} 1 & 0 \\ 0 & 0 \ea \right] [a])
\nonumber \\ \nonumber \\
&& +e^2 A_\mu A^\mu
([a]^\dagger \left[ \ba{cc} 1 & 0 \\ 0 & 0 \ea \right] [a]
+ [b]^\dagger \left[ \ba{cc} 1 & 0 \\ 0 & 0 \ea \right] [b])
\nonumber \\ \nonumber \\
&&-\frac{1}{4}F^{\mu \nu}F_{\mu \nu} - 
\frac{1}{2}\bar{\lambda} \not \! \partial \lambda + \frac{1}{2} D^2, \nonumber 
\eea
and the problem of ``interpreting'' auxiliary fields is therefore side-stepped.

We shall denote the propagators or proper vertices involving $[a]$ or $[b]$ by
enclosing them in square brackets to distinguish them from the propagators 
or vertices of the single component fields $a,b,f$ and $g$. Thus the $[a]$ and
$[b]$ propagators are
\be
[D(p^2)] \equiv
\left[ \ba{cc} \langle a^* a\rangle & \langle a^* f\rangle \\ 
\langle f^* a\rangle & \langle f^* f\rangle \ea \right]
= \left[ \ba{cc} \langle b^* b\rangle & \langle b^* g\rangle \\ 
\langle g^* b\rangle & \langle g^* g\rangle \ea \right];
\ee
their photon interaction is
\be
[\Gamma_{(a,b)^* A_\mu (a,b)}](p,q) \equiv
\left[ \ba{cc} \Gamma_{(a,b)^* A_\mu (a,b)}(p,q) 
& \Gamma_{(a,b)^* A_\mu (f,g)}(p,q) \\
\Gamma_{(f,g)^* A_\mu (a,b)}(p,q) 
& \Gamma_{(f,g)^* A_\mu (f,g)}(p,q) \ea \right];
\ee
the photino interactions are
\be
[\Gamma_{\lambdabar (a,b)^* \psi}](p,q) \equiv
[ \ba{cc} \Gamma_{\lambdabar (a,b)^* \psi}(p,q) &
\Gamma_{\lambdabar (f,g)^* \psi}(p,q) \ea ],
\ee
and
\be
[\Gamma_{\psibar (a,b) \lambda}](p,q) \equiv
\left[ \ba{c} \Gamma_{\psibar (a,b) \lambda}(p,q) \\
\Gamma_{\psibar (f,g) \lambda}(p,q)  \ea \right];
\ee
and their $D$ interactions are
\be
[\Gamma_{(a,b)^* D (b,a)}](p,q) \equiv
\left[ \ba{cc} \Gamma_{(a,b)^* D (b,a)}(p,q) 
& \Gamma_{(a,b)^* A_\mu (g,f)}(p,q) \\
\Gamma_{(f,g)^* A_\mu (b,a)}(p,q) 
& \Gamma_{(f,g)^* A_\mu (g,f)}(p,q) \ea \right].
\ee
One readily checks that 
Eqs.(\ref{twoeleven}) to (\ref{twothirteen}) are consistent with
\be
[\Gamma_{(a,b)^* (a,b)}](p) \equiv
\left[ \ba{cc} \Gamma_{(a,b)^* (a,b)}(p) 
& \Gamma_{(a,b)^* (f,g)}(p) \\
\Gamma_{(f,g)^* (a,b)}(p) & \Gamma_{(f,g)^* (f,g)}(p) \ea \right]
= [D(p^2)]^{-1}.
\ee

With the Lagrangian in its familiar form and our notation established it is
a simple matter to write down the DSE for the electron in SQED, namely
\bea
\lefteqn{S^{-1}(p) - S_{bare}^{-1}(p)} \\
&=&-\int \frac{d^4 p}{(2\pi)^4}\{ D_{\mu \nu}(p-q) \gamma^\mu S(q) 
\Gamma_{\psibar A_\mu \psi}^\nu (q,p) +
D_\lambda (p-q) [ \ba{cc} 1 & 0 \ea ] [D(q)] [\Gamma_{\lambdabar a^* \psi}](q,p)\}, \nonumber
\eea
where $D_{\mu \nu}$ is the photon propagator, and $D_\lambda$ the photino 
propagator.

\setcounter{equation}{0}
\section{Supersymmetric Vertex Ward Identities}
\label{SWI}

Before we can find the vertices to substitute into the SDSE, we need the
SWIs which constrain them. These are found by taking functional 
\begin{table}[h]
\caption{Each SWI is derived from a functional derivative of 
$\delta_S \Gamma =0$. The functional derivative leading to each 
SWI (indicated by its equation number) is given in this table.\label{fnctl}}
\vspace{0.2cm}
\begin{center}
\footnotesize
\begin{tabular}{|c|c||c|c|} 
\hline
\raisebox{0pt}[13pt][7pt]{Functional Derivative of
$\delta_S \Gamma =0$} &
\raisebox{0pt}[13pt][7pt]{SWI} &
\raisebox{0pt}[13pt][7pt]{Functional Derivative of
$\delta_S \Gamma =0$} &
\raisebox{0pt}[13pt][7pt]{SWI}\\
\hline
\raisebox{0pt}[13pt][7pt]{$\delta^3 \Gamma/
(\delta a(y) \delta a^* (x) \delta \lambdabar (z))$}&
\raisebox{0pt}[13pt][7pt]{\ref{ephotinoa}} &
\raisebox{0pt}[13pt][7pt]{$\delta^3 \Gamma/
(\delta \psi(y) \delta D(z) \delta a^*(x))$} &
\raisebox{0pt}[13pt][7pt]{\ref{aDpsi}}\\
\hline
\raisebox{0pt}[13pt][7pt]{$\delta^3 \Gamma/
(\delta b(y) \delta b^* (x) \delta \lambdabar (z))$} &
\raisebox{0pt}[13pt][7pt]{\ref{ephotinob}} &
\raisebox{0pt}[13pt][7pt]{$\delta^3 \Gamma/
(\delta \psi(y) \delta D(z) \delta b^*(x))$} &
\raisebox{0pt}[13pt][7pt]{\ref{bDpsi}}\\
\hline
\raisebox{0pt}[13pt][7pt]{$\delta^3 \Gamma/
(\delta f(y) \delta a^*(x) \delta \lambdabar(z))$} &
\raisebox{0pt}[13pt][7pt]{\ref{ephotinof}} &
\raisebox{0pt}[13pt][7pt]{$\delta^3 \Gamma/
(\delta \psi(y) \delta D(z) \delta f^*(x))$} &
\raisebox{0pt}[13pt][7pt]{\ref{fDpsi}}\\
\hline
\raisebox{0pt}[13pt][7pt]{$\delta^3 \Gamma/
(\delta g(y) \delta b^*(x) \delta \lambdabar(z))$} &
\raisebox{0pt}[13pt][7pt]{\ref{ephotinog}} &
\raisebox{0pt}[13pt][7pt]{$\delta^3 \Gamma/
(\delta \psi(y) \delta D(z) \delta g^*(x))$} &
\raisebox{0pt}[13pt][7pt]{\ref{gDpsi}}\\
\hline
\raisebox{0pt}[13pt][7pt]{$\delta^3 \Gamma/
(\delta f(y) \delta f^*(x) \delta \lambdabar(z))$} &
\raisebox{0pt}[13pt][7pt]{\ref{efpsi}} &
\raisebox{0pt}[13pt][7pt]{$\delta^3 \Gamma/
(\delta b(y) \delta D(z) \delta a^*(x))$} &
\raisebox{0pt}[13pt][7pt]{\ref{alambdab}}\\
\hline
\raisebox{0pt}[13pt][7pt]{$\delta^3 \Gamma/
(\delta g(y) \delta g^*(x) \delta \lambdabar(z))$} &
\raisebox{0pt}[13pt][7pt]{\ref{egpsi}} &
\raisebox{0pt}[13pt][7pt]{$\delta^3 \Gamma/
(\delta a(y) \delta \lambda (z) \delta b^*(x))$} &
\raisebox{0pt}[13pt][7pt]{\ref{blambdaa}}\\
\hline
\raisebox{0pt}[13pt][7pt]{$\delta^3 \Gamma/
(\delta \psi(y) \delta A_\mu(z) \delta f^*(x))$} &
\raisebox{0pt}[13pt][7pt]{\ref{ephotone}} &
\raisebox{0pt}[13pt][7pt]{$\delta^3 \Gamma/
(\delta g(y) \delta \lambda (z) \delta a^*(x))$} &
\raisebox{0pt}[13pt][7pt]{\ref{alambdag}}\\
\hline
\raisebox{0pt}[13pt][7pt]{$\delta^3 \Gamma/
(\delta \psi(y) \delta A_\mu(z) \delta g^*(x))$} &
\raisebox{0pt}[13pt][7pt]{\ref{ephotone2}} &
\raisebox{0pt}[13pt][7pt]{$\delta^3 \Gamma/
(\delta f(y) \delta \lambda (z) \delta b^*(x))$} &
\raisebox{0pt}[13pt][7pt]{\ref{blambdaf}}\\
\hline
\raisebox{0pt}[13pt][7pt]{$\delta^3 \Gamma/
(\delta \psi(y) \delta A_\mu(z) \delta a^*(x))$} &
\raisebox{0pt}[13pt][7pt]{\ref{apsi}} &
\raisebox{0pt}[13pt][7pt]{$\delta^3 \Gamma
(\delta a(y) \delta \lambda (z) \delta g^*(x))$} &
\raisebox{0pt}[13pt][7pt]{\ref{glambdaa}}\\
\hline
\raisebox{0pt}[13pt][7pt]{$\delta^3 \Gamma/
(\delta \psi(y) \delta A_\mu(z) \delta b^*(x))$} &
\raisebox{0pt}[13pt][7pt]{\ref{bpsi}} &
\raisebox{0pt}[13pt][7pt]{$\delta^3 \Gamma/
(\delta b(y) \delta \lambda (z) \delta f^*(x))$} &
\raisebox{0pt}[13pt][7pt]{\ref{flambdab}}\\
\hline
\raisebox{0pt}[13pt][7pt]{$\delta^3 \Gamma/
(\delta \psi_\alpha (y) \delta \psibar^\beta (x) \delta \lambda_\kappa (z))$} &
\raisebox{0pt}[13pt][7pt]{\ref{indices}} &
\raisebox{0pt}[13pt][7pt]{$\delta^3 \Gamma/
(\delta g(y) \delta \lambda (z) \delta f^*(x))$} &
\raisebox{0pt}[13pt][7pt]{\ref{flambdag}}\\
\hline
\multicolumn{2}{c||}{}&
\raisebox{0pt}[13pt][7pt]{$\delta^3 \Gamma/
(\delta f(y) \delta \lambda (z) \delta g^*(x))$} &
\raisebox{0pt}[13pt][7pt]{\ref{glambdaf}}\\
\cline{3-4}
\end{tabular}
\end{center}
\end{table}
derivatives of $\delta_S \Gamma =0$ where $\Gamma$ is the effective action
and $\delta_S$ is defined in Eqn.(\ref{super}). 
The functional derivatives of $\delta_S \Gamma =0$  corresponding to the
following SWIs are given in table \ref{fnctl}:

\bea 
\label{ephotinoa}
\lefteqn{\gamma_\mu \Gamma^\mu_{a^* A_\mu a}(p,q)} \\
&=& \Gamma_{\lambdabar a^* \psi}(p,q) \slash q + e(\Bp2 - \Bq2)
+ \Gamma_{\lambdabar a^* \psi}(-q,-p) \slash p, \nonumber \\
\label{ephotinob}
\lefteqn{\gamma_\mu \Gamma^\mu_{b^* A_\mu b}(p,q)} \\
&=& -i\Gamma_{\lambdabar b^* \psi}(p,q)\gamma_5 \slash q - e(\Bp2 - \Bq2)
- i\Gamma_{\lambdabar b^* \psi}(-q,-p)\gamma_5 \slash p, \nonumber
\eea
\bea
\label{ephotinof}
\gamma_\mu \Gamma^\mu_{f^* A_\mu a}(p,q) + e \Ap2
&=&  \Gamma_{\lambdabar a^* \psi}(-q,-p) 
+ \Gamma_{\lambdabar f^* \psi}(p,q)\slash q, \\
\label{ephotinog}
\gamma_\mu \Gamma^\mu_{g^* A_\mu b}(p,q) -e \Ap2 
&=&  i\Gamma_{\lambdabar b^* \psi}(-q,-p) \gamma_5 
+ i\Gamma_{\lambdabar g^* \psi}(p,q) \slash q \gamma_5,
\eea
\bea
\label{efpsi}
\gamma_\mu \Gamma^\mu_{f^* A_\mu f}(p,q)
&=&  \Gamma_{\lambdabar f^* \psi}(-q,-p) 
- \Gamma_{\lambdabar f^* \psi}(p,q), \\
\label{egpsi}
\gamma_\mu \Gamma^\mu_{g^* A_\mu g}(p,q)
&=& i\Gamma_{\lambdabar g^* \psi}(-q,-p)\gamma_5 
- i\Gamma_{\lambdabar g^* \psi}(p,q) \gamma_5, 
\eea
\bea 
\label{ephotone}
\lefteqn{i\sigma^{\mu \nu}(p-q)_\nu \Gamma_{\lambdabar f^* \psi}(p,q)} \\
& = & \Gamma^\mu_{\psibar A_\mu \psi}(p,q) 
-i\slash q \Gamma^\mu_{f^* A_\mu f}(p,q)
+i\Gamma^\mu_{f^* A_\mu a}(p,q) - ie\gamma^\mu \Ap2,  \nonumber \\
\label{ephotone2}
\lefteqn{i\sigma^{\mu \nu}(p-q)_\nu \Gamma_{\lambdabar g^* \psi}(p,q)} \\
& = & i\gamma_5 \Gamma^\mu_{\psibar A_\mu \psi}(p,q)
+\gamma_5 \slash q \Gamma^\mu_{g^* A_\mu g}(p,q)
-\gamma_5 \Gamma^\mu_{g^* A_\mu b}(p,q) + e\gamma_5 \gamma^\mu \Ap2, 
\nonumber 
\eea
\bea \label{apsi}
\lefteqn{i\sigma^{\mu \nu}(p-q)_\nu \Gamma_{\lambdabar a^* \psi}(p,q)} \\ 
& = &
i\Gamma^\mu_{a^* A_\mu a}(p,q)
- i\slash q \Gamma^\mu_{a^* A_\mu f}(p,q) -e\gamma^\mu S^{-1}(p) \nonumber
-\slash p \Gamma^\mu_{\psibar A_\mu \psi}(p,q) \\
&& + ie\gamma^\mu \Bp2, \nonumber \\
\lefteqn{i\sigma^{\mu \nu}(p-q)_\nu \Gamma_{\lambdabar b^* \psi}(p,q)} 
\label{bpsi} \\ & = &
-\gamma_5 \Gamma^\mu_{b^* A_\mu b}(p,q) 
+ \gamma_5 \slash q \Gamma^\mu_{b^* A_\mu g}(p,q) \nonumber
-i\gamma_5 e\gamma^\mu S^{-1}(p) \\
&& - i\gamma_5 \slash p \Gamma^\mu_{\psibar A_\mu \psi}(p,q)
- e\gamma_5 \gamma^\mu \Bp2. \nonumber
\eea
It follows from both (\ref{apsi}) and (\ref{bpsi}) that the rainbow
approximation, that is, dressed vertices replaced by bare vertices,
violates SUSY in the same way that it violates $U(1)$ gauge invariance.

From
\bea \label{indices}
0 &=& -i(\slash q)_\sigma^{\; \; \alpha}
(\Gamma_{\psibar f \lambda}(p,q))_\beta^{\; \; \kappa}
+ (\gamma_5 \slash q)_\sigma^{\; \; \alpha}
(\Gamma_{\psibar g \lambda}(p,q))_\beta^{\; \; \kappa} \\
&& - i(\slash p C)_{\beta \sigma}
\frac{\delta^2}{\delta \psi_\alpha(q) \delta f^*(p)}
(\frac{\delta \Gamma}{\delta \lambda(p-q)} C^{-1})^\kappa \nonumber \\
&& -(\gamma_5\slash p C)_{\beta \sigma}
\frac{\delta^2}{\delta \psi_\alpha(q) \delta g^*(p)}
(\frac{\delta \Gamma}{\delta \lambda(p-q)} C^{-1})^\kappa \nonumber \\
&& - (\gamma_5 (\slash p - \slash q))_\sigma^{\; \; \kappa}
(\Gamma_{\psibar D \psi}(p,q))_\beta^{\; \; \alpha}, \nonumber
\eea
where $C$ is the charge conjugation matrix, we obtain
\bea
0 & = &
(\slash p - \slash q)\gamma_5 \Tr(\Gamma_{\psibar D \psi}(p,q))
+ \gamma_\mu \Tr(\Gamma^\mu_{\psibar A_\mu \psi}(p,q))
+ i\Gamma_{\psibar a \lambda}(p,q) \nonumber \\
&&- \gamma_5 \Gamma_{\psibar b \lambda}(p,q) 
 - i\Gamma_{\psibar a \lambda\psi}(-q,-p)
+ \gamma_5 \Gamma_{\psibar b \lambda}(-q,-p) \\
&& - i\slash q \Gamma_{\psibar f \lambda}(p,q) 
+ \gamma_5 \slash q \Gamma_{\psibar g \lambda}(p,q) 
- i\slash p \Gamma_{\psibar f \lambda}(-q,-p) \nonumber \\
&& + \gamma_5 \slash p \Gamma_{\psibar g \lambda}(-q,-p), \nonumber
\eea
by setting $\beta = \alpha$ and summing, and
\bea \label{eDpsi}
0 & = & i\Tr(\Gamma_{\psibar a \lambda} (p,q))
- \gamma_5 \Tr(\Gamma_{\psibar b \lambda} (p,q)) 
- i\slash q \Tr(\Gamma_{\psibar f \lambda} (p,q))  \\
&& + \gamma_5 \slash q \Tr(\Gamma_{\psibar g \lambda} (p,q))
- i\Gamma_{\lambdabar a^* \psi}(p,q)
+ \gamma_5 \Gamma_{\lambdabar b^* \psi}(p,q)
- i\slash p \Gamma_{\lambdabar f^* \psi}(p,q) \nonumber \\
&& - \slash p \gamma_5 \Gamma_{\lambdabar g^* \psi}(p,q)) 
+ \gamma_\mu \Gamma^\mu_{\psibar A_\mu \psi} (p,q) 
- \gamma_5 (\slash p - \slash q) \Gamma_{\psibar D \psi} (p,q), \nonumber 
\eea
by setting $\beta = \kappa$ and summing.

Finally there are the SWIs governing the vertices of the $D$ particle;
\bea \label{aDpsi}
\lefteqn{i\gamma_5 \Gamma_{\lambdabar a^* \psi}(p,q)} \\
& = & \slash p \Gamma_{\psibar D \psi} (p,q) 
+ \gamma_5 \Gamma_{a^* D b}(p,q)
- \gamma_5 \slash q \Gamma_{a^* D g}(p,q), \nonumber \\
\label{bDpsi} \lefteqn{i\gamma_5 \Gamma_{\lambdabar b^* \psi}(p,q)} \\
& = & i\gamma_5 \slash p \Gamma_{\psibar D \psi} (p,q) 
- i\Gamma_{b^* D a}(p,q) + i\slash q \Gamma_{b^* D f}(p,q), \nonumber 
\eea
\bea \label{fDpsi}
\lefteqn{\gamma_5 \Gamma_{f^* D b}(p,q)} \\
& = & i\gamma_5 \Gamma_{\lambdabar f^* \psi}(p,q)
 + \gamma_5 \slash q \Gamma_{f^* D g}(p,q) 
+ \Gamma_{\psibar D \psi} (p,q), \nonumber \\
\lefteqn{\gamma_5 \Gamma_{g^* D a}(p,q)} \label{gDpsi} \\
& = & - \Gamma_{\lambdabar g^* \psi}(p,q)
+ \gamma_5 \slash q \Gamma_{g^* D f}(p,q)
- \Gamma_{\psibar D \psi} (p,q), \nonumber
\eea
\bea \label{alambdab}
\lefteqn{\gamma_5(\slash p - \slash q) \Gamma_{a^* D b}(p,q)} \\
& = & \Gamma_{\lambdabar b^* \psi}(-q,-p) \slash p
+ i\Gamma_{\lambdabar a^* \psi}(p,q) \slash q \gamma_5 
+ ie\gamma_5 (\Bp2 - \Bq2), \nonumber \\
\lefteqn{\gamma_5(\slash p - \slash q) \Gamma_{b^* D a}(p,q)} 
\label{blambdaa} \\
& = & i\Gamma_{\lambdabar a^* \psi}(-q,-p) \slash p \gamma_5
+ \Gamma_{\lambdabar b^* \psi}(p,q) \slash q
+ ie\gamma_5 (\Bp2 - \Bq2), \nonumber
\eea
\bea \label{alambdag}
\lefteqn{\gamma_5(\slash p - \slash q) \Gamma_{a^* D g}(p,q)} \\ 
& = &\Gamma_{\lambdabar g^* \psi}(-q,-p) \slash p
- i\Gamma_{\lambdabar a^* \psi}(p,q) \gamma_5 + ie\gamma_5 \Aq2, \nonumber \\ 
\lefteqn{\gamma_5(\slash p - \slash q) \Gamma_{b^* D f}(p,q)} 
\label{blambdaf} \\
& = &i\Gamma_{\lambdabar f^* \psi}(-q,-p) \slash p \gamma_5 
- \Gamma_{\lambdabar b^* \psi}(p,q) + ie\gamma_5 \Aq2, \nonumber \\
\lefteqn{\gamma_5(\slash p - \slash q) \Gamma_{g^* D a}(p,q)} 
\label{glambdaa} \\ 
& = &\Gamma_{\lambdabar g^* \psi}(p,q) \slash q
+ i\Gamma_{\lambdabar a^* \psi}(-q,-p) \gamma_5 - ie\gamma_5 \Ap2, \nonumber \\
\lefteqn{\gamma_5(\slash p - \slash q) \Gamma_{f^* D b}(p,q)} 
\label{flambdab} \\ 
& = &i\Gamma_{\lambdabar f^* \psi}(p,q) \slash q \gamma_5 
+ \Gamma_{\lambdabar b^* \psi}(-q,-p) - ie\gamma_5 \Ap2, \nonumber
\eea
\bea
\gamma_5(\slash p - \slash q) \Gamma_{f^* D g}(p,q) 
& = & \Gamma_{\lambdabar g^* \psi}(-q,-p) \label{flambdag}
- i\Gamma_{\lambdabar f^* \psi}(p,q) \gamma_5, \\
\gamma_5(\slash p - \slash q) \Gamma_{g^* D f}(p,q) 
& = & i\Gamma_{\lambdabar f^* \psi}(-q,-p) \gamma_5
- \Gamma_{\lambdabar g^* \psi}(p,q) \label{glambdaf}.
\eea

These make up the entire set of SWIs containing only three-or-fewer point
proper functions, modulo charge conjugation. A suitable vertex ansatz must
also be consistent with the WTIs;
\bea \label{WTI}
(p-q)_\mu [\Gamma_{(a.b)^* A_\mu (a,b)}]^\mu (p,q) &=&
e[\Gamma_{(a,b)^* (a,b)}] (p) - e[\Gamma_{(a,b)^* (a,b)}] (q), \\
(p-q)_\mu \Gamma^\mu_{\psibar A_\mu \psi}(p,q) &=&
eS^{-1}(p) - eS^{-1}(q).
\eea
We also have from charge conjugation invariance that
\be \ba{l} \label{conjugate}
\mbox{[$\Gamma_{\psibar (a,b) \lambda}$]}(p,q) = 
- C [\Gamma_{\lambdabar (a^*,b^*) \psi}](-q,-p)^T C^{-1}, \\
\mbox{[$\Gamma_{(a^*,b^*) D (b,a)}$]}(p,q) = 
- [\Gamma_{(b^*,a^*) D (a,b)}](-q,-p).
\ea \ee

\setcounter{equation}{0} 
\section{Solution to SWIs and WTIs in SQED} \label{threepoint}
Below is a solution for the SWIs and WTIs. It is the
most general set of vertices consistent with both the WTIs and the SWIs and
free of kinematic singularities if one assumes charge conjugation invariance
and
\be \label{condition}
[\Gamma_{a^* A_\mu a}]^\mu (p,q) = [\Gamma_{b^* A_\mu b}]^\mu (p,q).
\ee
Proof of this is presented in Appendix \ref{proof}. The assumption of 
Eqn.(\ref{condition}) is true to all orders in perturbation
theory, and any nonperturbative violations of this assumption are 
restricted by the WTIs to lie completely within their transverse components.

Our general solution is as follows: \newline
The scalar-photon vertices are
\bea
\lefteqn{\Gamma^\mu_{a^* A_\mu a}(p,q) 
= \Gamma^\mu_{b^* A_\mu b}(p,q)} \label{onea} \\
&=& \frac{e}{p^2 - q^2}(p^2 \Ap2 - q^2 \Aq2)(p+q)^\mu
+ [p^\mu (q^2 - p\cdot q) + q^\mu (p^2 - p\cdot q)]
T_{aa}(p^2,q^2),p\cdot q), \nonumber \\
\lefteqn{\Gamma^\mu_{a^* A_\mu f}(p,q) = \Gamma^\mu_{b^* A_\mu g}(p,q)
= \Gamma^\mu_{f^* A_\mu a}(p,q) 
= \Gamma^\mu_{g^* A_\mu b}(p,q)} \label{oneb} \\
&=& \frac{-e}{p^2 - q^2}(\Bp2 - \Bq2)(p+q)^\mu
+ [p^\mu (q^2 - p\cdot q) + q^\mu (p^2 - p\cdot q)]
T_{af}(p^2,q^2,p\cdot q) \nonumber, \\
\lefteqn{\Gamma^\mu_{f^* A_\mu f}(p,q) 
= \Gamma^\mu_{g^* A_\mu g}(p,q)} \label{onec} \\
&=& \frac{e}{p^2 - q^2}(\Ap2 - \Aq2)(p+q)^\mu
+ [p^\mu (q^2 - p\cdot q) + q^\mu (p^2 - p\cdot q)]
T_{ff}(p^2,q^2,p\cdot q), \nonumber
\eea
where the three functions 
$T_{aa}(p^2,q^2,p\cdot q),T_{af}(p^2,q^2,p\cdot q)$ and
$T_{ff}(p^2,q^2,p\cdot q)$, each satisfying
$T(p^2,q^2,p\cdot q) = T(q^2,p^2,p\cdot q)$, are free of
kinematic singularities and represent the only
degrees of freedom inherent in the solution. The forms (\ref{onea}) to
(\ref{onec}) are equivalent to that given by Ball and Chiu \cite{ball} in the
context of non SUSY scalar QED.
The photino vertices are
\bea \label{thirtysix} 
\Gamma_{\lambdabar a^* \psi}(p,q) &=& \frac{e}{p^2 - q^2}(p^2 \Ap2 - q^2 \Aq2)
+ \frac{e}{p^2 - q^2}(\Bp2 - \Bq2)\slash q \nonumber \\
&& + \half e (p^2 - \slash q \slash p)T_{aa}(p^2,q^2,p\cdot q) \\
&& + \half e p^2(q^2 - \slash p \slash q)T_{ff}(p^2,q^2,p\cdot q) 
\nonumber \\
&&+ \half e [\slash p (p^2 - q^2) - 2\slash q (p^2 - p\cdot q)]
T_{af}(p^2,q^2,p\cdot q), \nonumber
\eea

and
\bea \label{thirtyfive} 
\Gamma_{\lambdabar f^* \psi}(p,q) &=& \frac{-e}{p^2 - q^2}(\Ap2 - \Aq2)\slash q
-\frac{e}{p^2 - q^2}(\Bp2 - \Bq2) \nonumber \\
&& +\half e(\slash p - \slash q)T_{aa}(p^2,q^2,p\cdot q) \\
&& +  \half e(p-q)^2 T_{af}(p^2,q^2,p\cdot q) \nonumber \\
&& - \half e\slash q(p^2 - \slash p \slash q)T_{ff}(p^2,q^2,p\cdot q).
\nonumber
\eea
The electron-photon vertex must be restricted at least to the form given by
Ball and Chiu\cite{ball} for non SUSY QED. For the SUSY case we find
\bea
\Gamma^\mu_{\psibar A_\mu \psi}(p,q) &=&
\Gamma^\mu_{BC}(p,q) + \frac{ie}{p^2 - q^2}(\Ap2 - \Aq2)
	[\half T_3^\mu - T_8^\mu] \nonumber \\
&&		- \frac{ie}{p^2 - q^2}(\Bp2 - \Bq2)T_5^\mu
		+ \half ie T_{aa}(p^2,q^2,p\cdot q) T_3^\mu \nonumber \\
&& +ie T_{af}(p^2,q^2,p\cdot q)
	[\half (p-q)^2 T_5^\mu - T_1^\mu] \\
&& + \half ie T_{ff}(p^2,q^2,p\cdot q)
	[T_2^\mu - p\cdot q T_3^\mu - (p-q)^2 T_8^\mu], \nonumber
\eea
where 
\bea
\Gamma^\mu_{BC}(p,q) &=& \half \frac{ie}{p^2 - q^2}(\slash p + \slash q)
(A(p^2) - A(q^2))(p+q)^\mu \\
&& + ie\half (A(p^2) + A(q^2)) \gamma^\mu
+ \frac{ie}{p^2 - q^2}(\Bp2 - \Bq2)(p+q)^\mu, \nonumber
\eea
\bea \label{transverse}
T_1^\mu &=& p^\mu(q^2 - p\cdot q) + q^\mu (p^2 - p\cdot q), \\
T_2^\mu &=& (\slash p + \slash q)T_1^\mu, \\
T_3^\mu &=& \gamma^\mu(p-q)^2 - (\slash p - \slash q)(p-q)^\mu], \\
T_5^\mu &=& \sigma^{\mu \nu}(p-q)_\nu, \\
T_8^\mu &=& \half(\slash p \slash q \gamma^\mu - \gamma^\mu \slash q \slash p).
\eea

Finally there are the vertices for the $D$-boson, namely,
\bea
\lefteqn{\Gamma_{a^* D b} (p,q) = -\Gamma_{b^* D a} (p,q)} \\
& = & \frac{ie}{p^2 - q^2}(p^2 \Ap2 - q^2 \Aq2) 
- iep\cdot q T_{a^* a}(p^2,q^2,p\cdot q) \nonumber \\
&& + \half ie p^2 q^2 T_{ff}(p^2,q^2,p\cdot q), \nonumber \\ 
\nonumber \\
\lefteqn{\Gamma_{f^* D g} (p,q) = -\Gamma_{g^* D f} (p,q)} \\
&=&\frac{ie}{p^2 - q^2}(\Ap2 - \Aq2) 
+ ie T_{a^* a}(p^2,q^2,p\cdot q) \nonumber \\
&&- ie p\cdot q T_{f^* f}(p^2,q^2,p\cdot q),
\nonumber
\eea
\bea
\Gamma_{g^* D a} (p,q)&=& \frac{ie}{p^2 - q^2}(\Bp2 - \Bq2) \\
&&-ie (q^2 - p\cdot q) T_{af}(p^2,q^2,p\cdot q), \nonumber \\
\Gamma_{a^* D g} (p,q) &=& \frac{-ie}{p^2 - q^2}(\Bp2 - \Bq2) \\
&&+ie (p^2 - p\cdot q) T_{af}(p^2,q^2,p\cdot q), \nonumber \\
\Gamma_{f^* D b} (p,q) &=& \frac{-ie}{p^2 - q^2}(\Bp2 - \Bq2) \\
&&+ie (q^2 - p\cdot q) T_{af}(p^2,q^2,p\cdot q), \nonumber \\
\Gamma_{b^* D f} (p,q) &=& \frac{ie}{p^2 - q^2}(\Bp2 - \Bq2) \\
&&-ie (p^2 - p\cdot q) T_{af}(p^2,q^2,p\cdot q), \nonumber 
\eea
and
\bea
\Gamma_{\psibar D \psi}(p,q) &=& 
\half ie\gamma_5 [(p^2 - q^2)T_{af}(p^2,q^2,p\cdot q) \\
&& +(\slash p + \slash q)T_{a^* a}(p^2,q^2,p\cdot q) \nonumber \\
&& -(\slash q p^2 + \slash p q^2)T_{ff}(p^2,q^2,p\cdot q)]. \nonumber
\eea

\setcounter{equation}{0}
\section{Conclusion}

We have derived the three-point SWIs for SQED and found a solution, given in
sections \ref{twopoint} and \ref{threepoint},
which, under the reasonable assumptions of charge conjugation invariance
and symmetry between $[a]$ and $[b]$
with respect to their photon interaction, comprises the most general
set of vertices consistent with both the SWIs and WTIs and free of kinematic
singularities. They are, in fact, the SUSY equivalent of the 
Ball-Chiu vertex. These SUSY Ball-Chiu  vertices 
have only three degrees of freedom 
between them once the electron propagator is known, 
compared with non SUSY QED which has eight. The loss of degrees of freedom
occurs entirely within the electron-photon vertex. The scalar-photon vertices
remain unchanged from non SUSY scalar QED (with auxiliary fields).

We have given the form of the electron DSE. There is no need to consider also
the DSE for scalar partners since SWIs ensure that the propagators of all
chiral multiplet fields can be written in terms of the same two scalar 
functions $\Ap2$ and $\Bp2$ (See Sec.(\ref{twopoint})). Solving the DSE for
any chiral multiplet field can therefore be accomplished by projecting from
the electron DSE a pair of coupled integral equations for $\Ap2$ and $\Bp2$.

Numerical solutions of the analogous calculation in non SUSY QED 
\cite{HW95,HWR96,HSW97,RW94} and QED$_3$ \cite{BR91,CPW92} 
using the minimal Ball-Chiu and Curtis-Pennington \cite{CP91} vertex 
ans\"atze exist in the literature. The same task in SUSY is conceptually
similar and the presence of extra terms in the DSE is not expected to 
reduce its feasibility. Indeed such numerical work has
been done already in the rainbow approximation in SQED$_3$ \cite{me}.
The way now lies
open to transcend the rainbow approximation in the analysis of SQED and 
SQED$_3$ in the nonperturbative limit.
\appendix
\renewcommand{\theequation}{\Alph{section}.\arabic{equation}}
\setcounter{section}{0}
\setcounter{equation}{0}
\section{Appendix: Derivation of the Nonperturbative Vertices} \label{proof}
Below is a derivation of the most general form of the proper vertices
consistent with both the SWIs and the
WTIs. It is convenient to define the following notation:

The operator $\Omega$ performs the 
interchange $(p,q) \longleftrightarrow (-q,-p)$.

A function $F(p,q)$, invariant to $\Omega$, is written as $F(_(p,q_))$.
If $F(p,q)$ is a scalar function $F(p^2,q^2,p\cdot q)$ then it is
written as $F(_(p^2,q^2_),p\cdot q)$. \newline
Alternately, a function $G(p,q)$ which changes sign under $\Omega$
is written as $G(_[p,q_])$, or $G(_[p^2,q^2_],p\cdot q)$ if it is
scalar.

Eqs.(\ref{onea}, \ref{oneb}, \ref{onec}) 
follow, by the reasoning of Ball and Chiu \cite{ball}, 
from the WTI for $[a]$ and $[b]$ (See Eqn.(\ref{WTI})).

Substituting Eqn.(\ref{condition}) into Eqn.(\ref{bpsi}) and
comparing to Eqn.(\ref{apsi}) gives
\be \label{cndtna}
\Gamma_{\lambdabar b^* \psi}(p,q) = i\gamma_5 \Gamma_{\lambdabar a^* \psi}(p,q).
\ee
Similarly, from Eqs.(\ref{ephotone}, \ref{ephotone2}),
\be \label{cndtnf}
\Gamma_{\lambdabar g^* \psi}(p,q) = i\gamma_5 \Gamma_{\lambdabar f^* \psi}(p,q).
\ee

Any $\Gamma_{\lambdabar f^* \psi}(p,q)$ consistent with Eqn.(\ref{efpsi})
can be put in the general form
\bea \label{three}
\Gamma_{\lambdabar f^* \psi}(p,q)
&=& \frac{-e}{p^2 - q^2} (\Ap2 - \Aq2)\slash q + H(_(p,q_)) \\
&& - \half e[\slash p (q^2 - p\cdot q) + \slash q (p^2 - p\cdot q)]
T_{ff}(_(p^2,q^2_),p\cdot q). \nonumber
\eea
Using Eqn.(\ref{cndtna}) to equate Eqs.(\ref{fDpsi}, \ref{gDpsi}), we find
that
\bea
\Gamma_{f^* D b} (p,q) &=& - \Gamma_{g^* D a} (p,q), \label{fivea} \\
\Gamma_{f^* D g} (p,q) &=& - \Gamma_{g^* D f} (p,q). \label{fiveb}
\eea
We obtain, by substituting Eqs.(\ref{cndtnf}, \ref{three}) into 
Eqn.(\ref{glambdaf}),
\bea \label{eight}
\lefteqn{\gamma_5(\slash p - \slash q)\Gamma_{g^* D f} (p^2,q^2,p\cdot q)}
\\
&=& \frac{-ie}{p^2 - q^2}(\Ap2 - \Aq2)\gamma_5 (\slash p - \slash q)
 + iH(_(p,q_))\gamma_5 -i\gamma_5 H(_(p,q_)). \nonumber
\eea
Dividing $H(_(p,q_))$ into its odd-numbered and even-numbered 
$\gamma$-matrix components,
$H^{\mbox{\scriptsize odd}}(_(p,q_))$ 
and $H^{\mbox{\scriptsize even}}(_(p,q_))$ respectively, we see from 
Eqn.(\ref{eight}) that $H^{\mbox{\scriptsize odd}}(_(p,q_))$ is of the form
\be \label{twelve}
H^{\mbox{\scriptsize odd}}(_(p,q_)) = (\slash p - \slash q)
\stackrel{\wedge}{H}(_(p^2,q^2_),p\cdot q),
\ee
due to its anti-commutation with $\gamma_5$ and its invariance under $\Omega$.
If we substitute Eqs.(\ref{cndtnf}, \ref{fivea}, \ref{conjugate}) into 
Eqn.(\ref{alambdag}) we get
\bea \label{star}
\lefteqn{\gamma_5(\slash p - \slash q)\Gamma_{b^* D f}(p^2,q^2,p\cdot q)} \\
&=& i \Gamma_{\lambdabar a^* \psi}(p,q) \gamma_5 - ie\gamma_5 \Aq2
-i\gamma_5 \Gamma_{\lambdabar f^* \psi}(-q,-p) \slash p, \nonumber
\eea
which, when added to Eqn.(\ref{blambdaf}), produces
\bea \label{fourteen}
\lefteqn{i\gamma_5 \Gamma_{\lambdabar a^* \psi}(p,q)
+ i \Gamma_{\lambdabar a^* \psi}(p,q) \gamma_5} \\
&=& 2ie\Aq2 + i\gamma_5 \Gamma_{\lambdabar f^* \psi}(-q,-p) \slash p
- i\Gamma_{\lambdabar f^* \psi}(-q,-p) \gamma_5 \slash p. \nonumber
\eea
Any $\Gamma_{\lambdabar a^* \psi}(p,q)$ consistent with 
Eqs.(\ref{three}, \ref{twelve}, \ref{fourteen}) must be of the form
\bea \label{fifteen}
\Gamma_{\lambdabar a^* \psi}(p,q) &=& \frac{e}{p^2 - q^2}(p^2 \Ap2 - q^2 \Aq2)
\\
&& + \half e [p^2 (q^2 - p\cdot q) + \slash q \slash p (p^2 - p\cdot q)]
T_{ff}(_(p^2,q^2_),p\cdot q) \nonumber \\
&& + (p^2 - \slash q \slash p)\stackrel{\wedge}{H}(_(p^2,q^2_),p\cdot q) 
+ \Gamma^{\mbox{\scriptsize odd}}_{\lambdabar a^* \psi}(p,q), \nonumber
\eea
where the superscript ``odd'' on the last term denotes that it is the
component of $\Gamma_{\lambdabar a^* \psi}(p,q)$ with only odd
numbers of $\gamma$-matrices.
$\Gamma^{\mbox{\scriptsize odd}}_{\lambdabar a^* \psi}(p,q)$ is unrestricted by
Eqn.(\ref{fourteen}) due to its anti-commutation with $\gamma_5$.

Substituting Eqs.(\ref{onea}, \ref{fifteen}) into
Eqn.(\ref{ephotinoa}) tells us that
\be \label{seventeen}
\stackrel{\wedge}{H}(_(p^2,q^2_),p\cdot q) = \half e
(T_{aa}(_(p^2,q^2_),p\cdot q)  - p\cdot q T_{ff}(_(p^2,q^2_),p\cdot q)).
\ee

The even $\gamma$-matrix component of $\Gamma_{\lambdabar a^* \psi}(p,q)$
is therefore
\bea \label{nineteen}
\Gamma^{\mbox{\scriptsize even}}_{\lambdabar a^* \psi}(p,q) &=&
\frac{e}{p^2 - q^2}(p^2 \Ap2 - q^2 \Aq2) \\
&& + \half e(p^2 - \slash q \slash p) T_{aa}(_(p^2,q^2_),p\cdot q) \nonumber \\
&& + \half e p^2 (q^2 - \slash p \slash q) T_{ff}(_(p^2,q^2_),p\cdot q),
\nonumber
\eea
and the odd $\gamma$-matrix component of $\Gamma_{\lambdabar f^* \psi}(p,q)$
is 
\bea \label{twentyone}
\Gamma^{\mbox{\scriptsize odd}}_{\lambdabar f^* \psi}(p,q) &=&
\frac{-e}{p^2 - q^2}(\Ap2 - \Aq2)\slash q \\
&& + \half e(\slash p - \slash q) T_{aa}(_(p^2,q^2_),p\cdot q) \nonumber \\
&& - \half e \slash q (p^2 - \slash p \slash q) T_{ff}(_(p^2,q^2_),p\cdot q).
\nonumber
\eea

It now remains to find 
$\Gamma^{\mbox{\scriptsize odd}}_{\lambdabar a^* \psi}(p,q)$
and $H^{\mbox{\scriptsize even}}(_(p,q_))$.
Subtracting Eqn.(\ref{star}) from Eqn.(\ref{blambdaf}) we get
\be \label{twentyfour}
(\slash p - \slash q)\Gamma_{b^* D f}(p^2,q^2,p\cdot q) =
-i\Gamma^{\mbox{\scriptsize odd}}_{\lambdabar a^* \psi}(p,q)
- iH^{\mbox{\scriptsize even}}(_(p,q_)) \slash p.
\ee
The result of substituting 
Eqs.(\ref{nineteen}, \ref{twentyone}) into Eqn.(\ref{ephotinof})
and operating with $\Omega$ is
\bea \label{twentyfive}
0 &=& \Gamma^{\mbox{\scriptsize odd}}_{\lambdabar a^* \psi}(p,q)
- H^{\mbox{\scriptsize even}}(_(p,q_)) \slash p \\
&& - \frac{e}{p^2 - q^2}(\Bp2 - \Bq2)(\slash p + \slash q) \nonumber \\
&& + e[\slash p (q^2 - p\cdot q) + \slash q (p^2 - p\cdot q)]
T_{af}(_(p^2,q^2_),p\cdot q). \nonumber
\eea
Adding Eqn.(\ref{twentyfive}) to $-i\times$\{Eqn.(\ref{twentyfour})\} produces
\bea \label{twentysix}
\lefteqn{-i(\slash p - \slash q)\Gamma_{b^* D f}(p^2,q^2,p\cdot q)} \\
&=& -2 H^{\mbox{\scriptsize even}}(_(p,q_)) \slash p
- \frac{e}{p^2 - q^2}(\Bp2 - \Bq2)(\slash p + \slash q) \nonumber \\
&& + e[\slash p (q^2 - p\cdot q) + \slash q (p^2 - p\cdot q)]
T_{af}(_(p^2,q^2_),p\cdot q). \nonumber
\eea

$H^{\mbox{\scriptsize even}}(_(p,q_))$ is of the general form,
\bea \label{twentyseven}
H^{\mbox{\scriptsize even}}(_(p,q_)) &=& 
H^{\mbox{\scriptsize scalar}}(_(p^2,q^2_),p\cdot q)
+ \gamma_5 H^5(_(p^2,q^2_),p\cdot q)\\ && 
+ \half(\slash p \slash q - \slash q \slash p) H^\sigma (_[p^2,q^2_],p\cdot q)
\nonumber \\ &&
+ \half \gamma_5 (\slash p \slash q - \slash q \slash p)
H^{5\sigma} (_[p^2,q^2_],p\cdot q). \nonumber
\eea
The symmetry properties of the scalar functions in Eqn.(\ref{twentyseven}) 
follow from the invariance of $H(_(p,q_))$ under $\Omega$.
Remembering that $\Gamma_{b^* D f}(p^2,q^2,p\cdot q)$ is scalar, and
substituting Eqn.(\ref{twentyseven}) into Eqn.(\ref{twentysix}), we find that
\be \label{thirty}
H^{5\sigma} (_[p^2,q^2_],p\cdot q) = 0 = H^5(_(p^2,q^2_),p\cdot q),
\ee
\be \label{thirtyone}
H^{\sigma} (_[p^2,q^2_],p\cdot q) = 0,
\ee
and
\bea \label{thirtytwo}
\lefteqn{H^{\mbox{\scriptsize scalar}}(_(p^2,q^2_),p\cdot q)} \\ 
&=& \half e (p-q)^2 T_{af}(_(p^2,q^2_),p\cdot q)
- \frac{e}{p^2 - q^2} (\Bp2 - \Bq2). \nonumber
\eea
Finally, substituting 
Eqs.(\ref{twentyseven}) to (\ref{thirtytwo}) into Eqn.(\ref{twentyfive}),
\bea \label{thirtyfour}
\Gamma^{\mbox{\scriptsize odd}}_{\lambdabar a^* \psi}(p,q) &=& 
\frac{e}{p^2 - q^2} (\Bp2 - \Bq2) \slash q \\
&& + \half e [\slash p (p^2 - q^2) - 2 \slash q (p^2 - p\cdot q)]
T_{af}(_(p^2,q^2_),p\cdot q). \nonumber
\eea

We now have the vertices $\Gamma_{\lambdabar f^* \psi}(p,q)$, given by
Eqn.(\ref{thirtyfive}), and $\Gamma_{\lambdabar a^* \psi}(p,q)$, found by
summing Eqs.(\ref{nineteen}) and (\ref{thirtyfour}) and given by
Eqn.(\ref{thirtysix}). $\Gamma^\mu_{\psibar A_\mu \psi}(p,q)$
is now determined by any one of the Eqs.(\ref{ephotone}) to
(\ref{bpsi}), the scalar $D$-vertices are given by the Eqs.(\ref{alambdab}) 
through to (\ref{glambdaf}), and the vertex
$\Gamma_{\psibar D \psi}(p,q)$ is given by any one of the Eqs.(\ref{aDpsi}) 
through to (\ref{gDpsi}). It is simple to verify that 
the solution presented in section \ref{threepoint} is not further constrained 
by the SWIs not used in this derivation.

\setcounter{equation}{0}
\paragraph{References}
 \end{document}